\documentclass{eas}
\usepackage{graphicx}
\begin{document}

\TitreGlobal{Mass Profiles and Shapes of Cosmological Structures}

%%-----------------------------
%%      the top matter
%%-----------------------------
\title{Polar Ring Galaxies and Warps}
\author{Combes, F.}
\address{Observatoire de Paris, LERMA, 61 AV. de l'Observatoire, F-75014, Paris, France}
\runningtitle{Polar Ring Galaxies and Warps}
\setcounter{page}{23}
% Keep this line, even if the page will be settled afterwards..
\index{Combes, F.}
% Repeat the authors here, this will help to make the final index

%
\begin{abstract} 
Polar ring galaxies, where matter is in equilibrium in perpendicular
orbits around spiral galaxies, are ideal objects to probe the 3D shapes
of dark matter halos. The conditions to constrain the halos are that the
perpendicular system does not strongly perturb the host galaxy, or that
it is possible to derive back its initial shape, knowing the formation
scenario of the polar ring. The formation mechanisms are reviewed:
mergers, tidal accretion, or gas accretion from cosmic filaments.
The Tully-Fisher diagram for polar rings reveals that the velocity
in the polar plane is higher than in the host plane, which can 
only be explained if the dark matter is oblate and flattened 
along the polar plane.  Only a few individual systems have
been studied in details, and 3D shapes of their haloes determined
by several methods. The high frequency of warps could be explained
by spontaneous bending instability, if the disks are sufficiently
self-gravitating, which 
can put constraints on the dark matter flattening. 
\end{abstract}

\maketitle
%
%%-----------------------------
%%      your text
%%-----------------------------
 
 \section{Polar rings formation scenarios}

Polar Ring Galaxies (PRG)  are composed of an early-type host,
surrounded by a gaseous and stellar  perpendicular ring
which has the characteristic of a late-type galaxy:
large HI amount, young stars, blue colors.
If the polar ring is in equilibrium, it could provide
clues on the 3D-shape of dark matter in the host galaxy.
 For that, the polar ring must not be too massive, 
in order to keep the initial gravitational potential unperturbed; or
in any case, we must know the formation mechanism of the polar rings, to
be able to generalise the results to normal haloes.
The probability to have a PRG has been estimated to $\sim$ 5\% 
from the observations (Whitmore et al 1990).

Several mechanisms have been proposed in the literature
to form polar ring galaxies, that can be gathered into two
kinds: merger between two galaxies, of which one at least is gas rich,
or mass accretion by the original host galaxy. 
The first kind can be a major merger between two perpendicularly oriented
systems, or the disruption of a small companion on a polar orbit (the debris
align then on the polar ring).
The second kind can  be either gas transfer between two galaxies,
in an hyperbolic passage, or gas accretion from cosmic filaments
in a polar ring by the host.
In all cases, the bulk of the stars in the polar plane form after
the event, from the gas settled  in the polar plane.

\subsection{Merger scenario}

The collision scenario has first been  simulated 
by Bekki (1997, 1998). Bournaud \& Combes (2003)
have explored the geometrical parameters of the collision
to determine the stability of the final system, and the 
actual probability to form a PRG from such a merger.
The mechanim is perfectly able to form polar rings,
and even double rings (see the ESO 474-G26 system in Figure 1, 
Reshetnikov et al 2005a). However the parameter space
is smaller than for accretion, so the majority of PRGs 
must not have been formed through mergers.

\begin{figure}[h]
   \centering
\begin{tabular}{cc}
   \includegraphics[width=4cm]{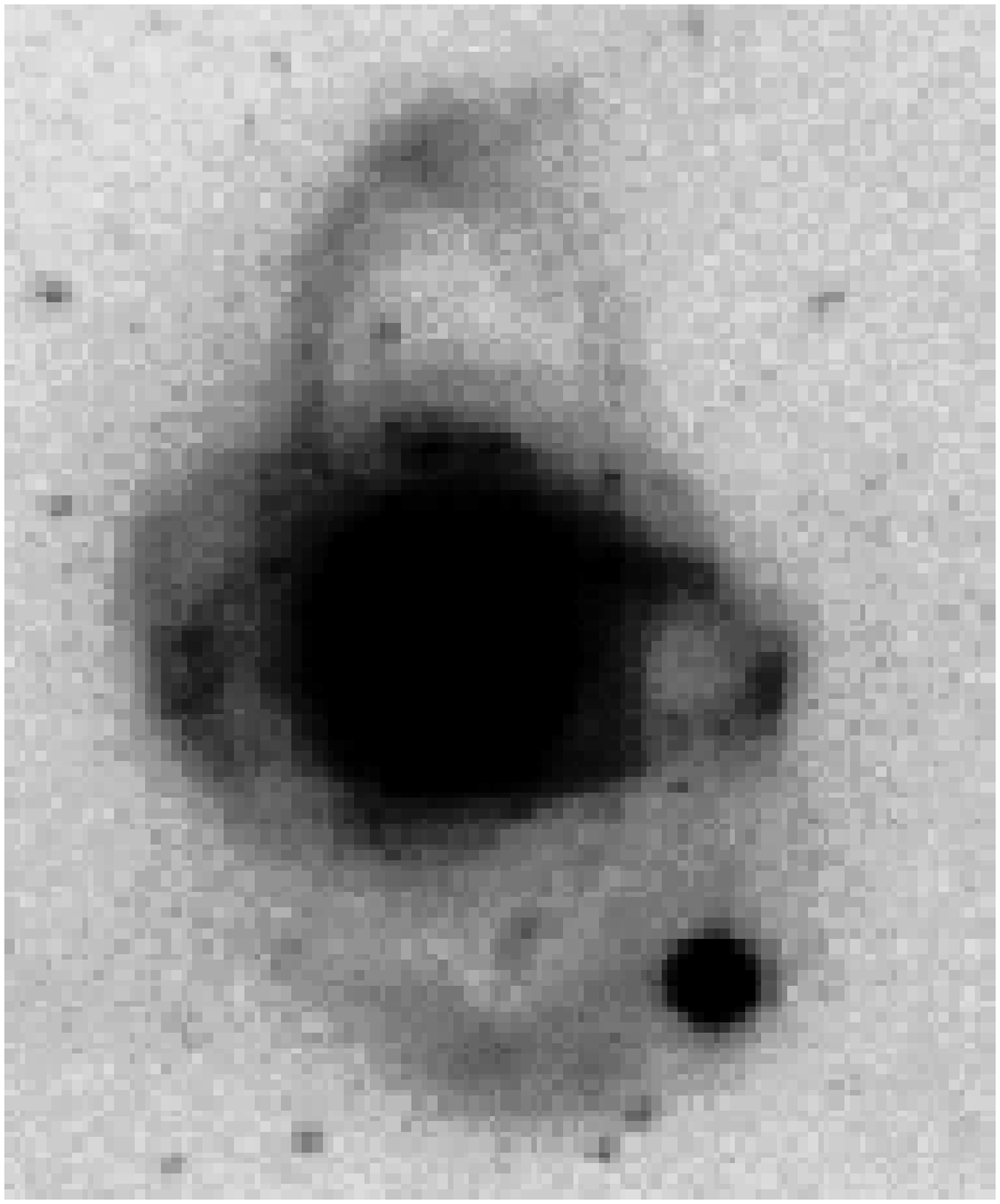}&
   \includegraphics[width=4cm]{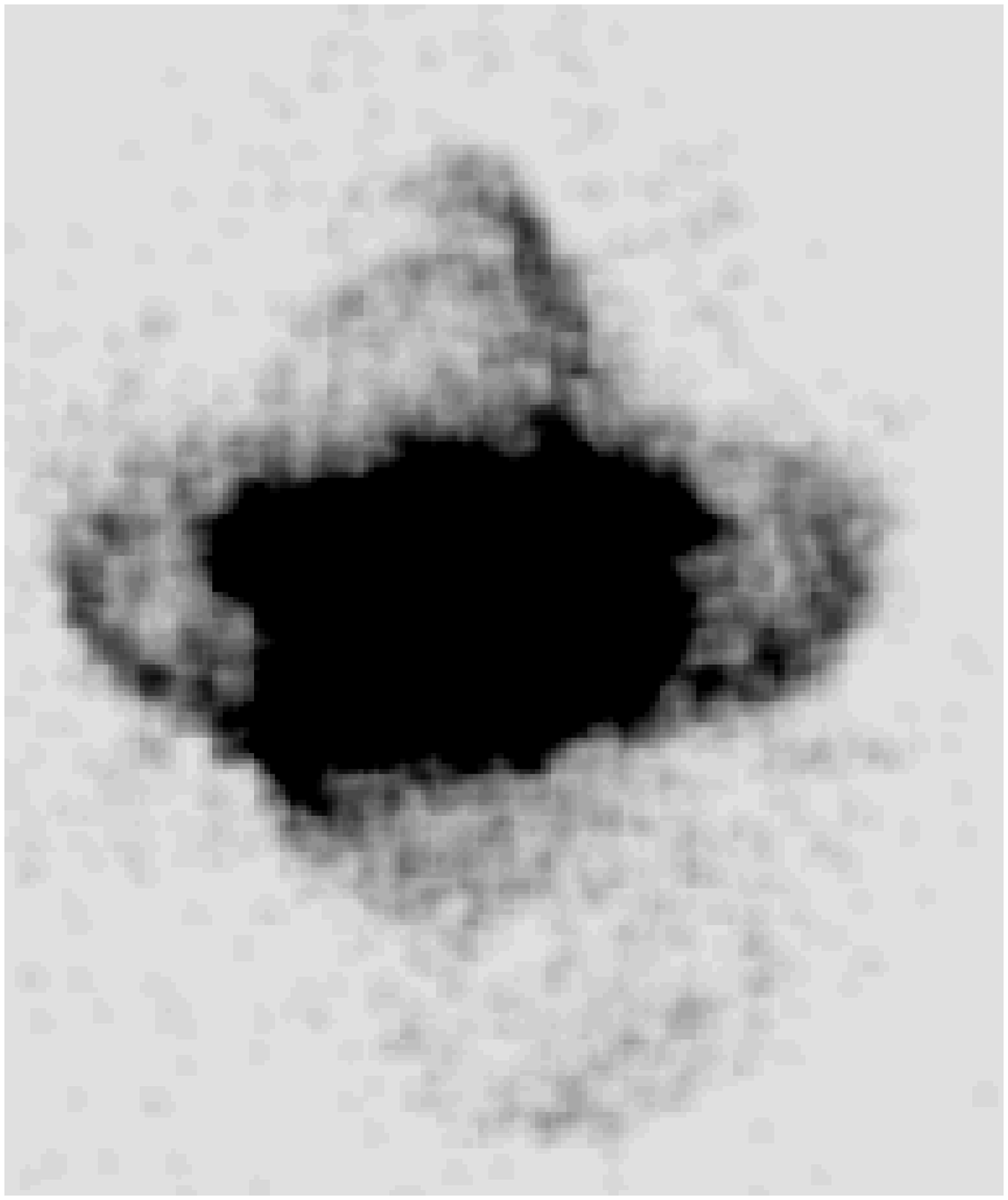}\\
\end{tabular}
      \caption{The double ringed galaxy ESO 474-G26:
{\bf Left:} The V-band image of the Galaxy 
{\bf Right:} The result of the simulated formation
of this system through a merger of two spiral galaxies of comparable 
masses, at t=700 Myr.  From Reshetnikov et al (2005a). }
       \label{fig1}
   \end{figure}

\subsection{Tidal accretion scenario}

The tidal accretion scenario, proposed
by Schweizer et al (1983) and first simulated by
Reshetnikov et al (1997), was long thought to form only
very light polar rings, since the donor galaxy exchange only
part of its mass, before leaving the interaction scene.
When the geometrical parameters are explored, they are
able to form quite inclined polar ring, such as the prototype
NGC 660, which makes the continuity with highly warped systems
(see also NGC 3718, Krips et al 2005). From the 
statistics of initial conditions, the accretion scenario
was estimated to be  3-5 times more likely
to form polar rings (Bournaud \& Combes 2003).
 The configuration of particular systems tends to orient
towards this accretion scenario, which provides the best fit,
(see for instance AM1934-563, Reshetnikov et al 2005b).

Very massive polar rings can form 
through accretion, with polar components of
comparable mass (or even larger) to the host component
The prototype NGC 4650A is a good case for accretion, since
no stellar halo is detected around the galaxy. Such a stellar halo
is expected in the merging scenario to come from the dispersed
stars from the destroyed satellite.
The polar component is quite massive,
8 10$^9$ M$_\odot$  of HI gas and 
4  10$^9$ M$_\odot$ in stars.

\subsection{Gas accretion from cosmic filaments}

The formation of polar rings could also occur
without any interaction or merger, through
gas infalling from cosmic filaments, with 
 inclined  angular momentum
(Katz \& Rix 1992, Semelin \& Combes 2005).
This scenario has been simulated by
Maccio \& Moore (2005): the gas is acquired from 
a filament of $\sim$ 1 Mpc long, and is essentially cold
gas accretion, with small impact parameter. This kind of cold gas
accretion might be the more realistic way by which galaxies
get their gas (e.g. Keres et al 2005).
  One case happens to be quite similar to the prototype NGC 4650A,
where the stars are essentially in the host and gas in the polar plane;
 several small companions are also found all along the filament.
Dark matter is quite round in the visible part
after the infall of gas.

\subsection{Late dark matter infall with resonance}

An alternative scenario, requiring no external gas accretion or merger for
PRG formation, has been proposed by Tremaine \& Yu (2000).
  It assumes that a disk galaxy lies in the symmetry plane of a
triaxial dark matter halo. The dark halo tumbles with a retrograde
pattern speed with respect to the disk. When this tumbling 
slows down, through dark matter infall, the stars in the disk could
get trapped in a vertical resonance with the halo. Indeed, a stellar
orbit slightly tilted with respect to the plane, precess at a slow retrograde 
rate, and when the resonance progressively reaches stars
at larger and larger radii, the stars are propulsed into the polar plane.   
With this mechanism, the polar ring would contain 
two equal-mass counter-rotating
stellar disks.
This mechanism also explains the splitting of the stars in two
counter-rotating streams in the disk, when the tumbling halo 
slows down from retrograde, to zero and then prograde
(Evans \& Collett 1994).

This mechanism however has not been confirmed in observations,
when velocity fields of the stars in the polar ring have been studied
(e.g. Swaters \& Rubin 2003).
The stellar kinematics in NGC 4650A for instance does
not reveal counter-rotation. 
A flat rotation curve in observed in the polar ring, which
confirms that it is more a full plane than a ring.
With the help of large telescopes, it is possible to have now
much more precise and accurate kinematical information
on the remote polar ring, as shown by the
new VLT-FORS2   data on NGC 4650A (Iodice et al 2005).
The detailed stellar kinematics on the major and minor axis
of the host galaxy reveal a flat velocity dispersion,
with a slight drop in the center, and will allow in the near future 
a better determination of the gravitational
potential and its 3D shape.

\section{Tully-Fisher diagram for PRGs}

Since a complete velocity field is difficult to obtain for many
PRGs, a single Tully-Fisher allows to gather more statistics
(Iodice et al 2003). The position of 15 polar rings revealed quite surprising
in such a plot, since the HI total widths of the gas in the polar plane
is clearly larger than expected from the normal spiral TF relation.
  Because of the superposition of the two perpendicular systems,
the orbits are expected to be non circular in the polar rings.
In general, and for obvious selection effects,
both components are seen nearly edge-on. Then the maximum
velocity in projection in the ring is coming from the top of the ring,
corresponding to the largest distance from the center of the galaxy. 
The observed velocity is then expected to be lower than circular, and
then lower than the maximal velocity observed in the host galaxy plane.
This is already the case in absence of dark matter, or if the dark matter
is spherical. If the dark halo is flattened and oblate, parallel to
the host disk, this effect will be accentuated, and the HI width in the
polar plane expected even lower. 

But the contrary is observed. Is this due to the fact that the host
component is also perturbed, and does not fall on the TF relation?
Figure 2 shows the TF diagram for the PRGs where a detailed 
velocity width have been obtained in the two perpendicular components.
It shows that the host galaxy obeys the normal TF relation, while there is 
a large difference of velocity between the two perpendicular planes,
the velocity being larger in the polar ring.

The only solution to the problem is to assume that the
dark matter halo is oblate and aligned with the polar ring itself.
In this case, the orbits in the host are then non-circular, and 
we observe the maximum velocity when the matter is farther 
from the center, and lower than circular.  In the polar plane,
the orbits will be circular, and larger than in the host plane.
It does not help to assume that the dark halo is prolate, keeping 
the orbits circular in the host. 

Is there a formation scenario able to explain that
most PRGs have their dark matter halo flattened along the polar disk?
With dissipationless dark matter, both merging and accretion scenarios
produce either spherical haloes, or halos flattened along the host.
However, if a large fraction of the dark matter around galaxies is dissipative
(dark gas), it will settle in the polar plane, and 
 it is possible to account for the flattening along the polar disk.

\begin{figure}[h]
   \centering
   \includegraphics[angle=270,width=9cm]{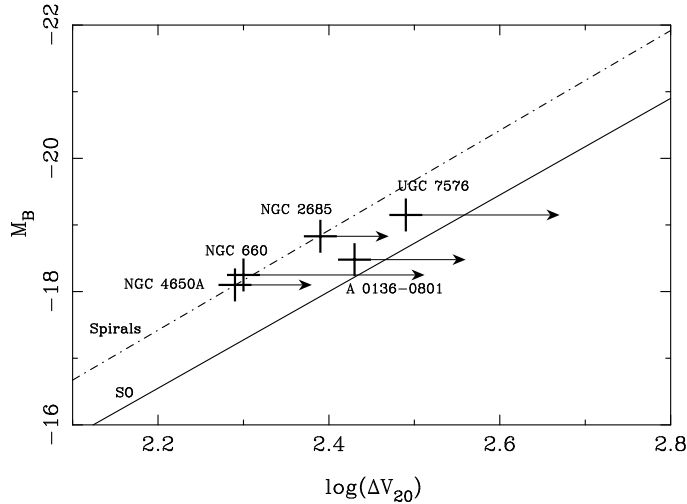}
      \caption{TF relation for spirals (dashed line), S0 galaxies (solid lines) and 5
PRGs, where the velocity curve is known in both perpendicular components.
Large crosses show the position of the central host galaxy  and
the arrow shows the offset in
$\log(\Delta V_{20})$ between the velocities in the host galaxy equatorial
plane and in the polar ring for each system.   From Iodice et al (2003).}
       \label{fig2}
   \end{figure}

\section{3D shapes of haloes}

\subsection{Axisymmetry}

Pure CDM simulations predict triaxial shapes for collapsed structures, so that the haloes are
not axisymmetric even in the plane of the baryonic galactic disk. 
The axisymmetry in galactic disks have been checked through
the orbits of the baryons, and in particular the HI gas, with low velocity dispersion. Of course,
 inclination effects have to be taken into account, as well as flaring, warps or other distortions,
 due to the spiral, bars or ring features in the galaxy disks.

The result of these investigations is that galaxies are actually
axisymmetric in their planes, with a very low upper limit for the excentricity: below 0.1 with the
 isophote shape versus
HI velocity widths method (Merrifield 2002), or even less than 0.045, when using near-infrared data
 to avoid extinction (Rix \& Zaritsky 1995). On special cases, the limit can be better, excentricity
 of the order of 0.012 in potential in the very regular early-type galaxy IC2006 with an HI ring
 (Franx et al 1994).
This axisymmetric shape of galactic haloes is confirmed by the low scatter observed for the
Tully-Fisher relation.

\subsection{Flattening}

As for the flattening, pure CDM haloes are predicted rather flattened 
in numerical simulations; they are  half oblate and half prolate,
with axis ratios of the order of c/a = 0.5, b/a = 0.7. It is interesting to note that the dark haloes
 are predicted more flattened then observed elliptical galaxies;
the distribution peaks at E5 (while elliptical galaxies
peak at E2 (c/a = 0.8)) cf Dubinski \& Carlberg (1991).

However, when the dissipative infall of gas is taken into account,
the dark matter haloes are concentrated, through adiabatic contraction, and are also 
forced to an oblate shape, the prediction now becomes in average: c/a = 0.5, b/a $>$ 0.7
(Dubinski 1994).
This depends a lot on the actual physics of baryons. When the cooling is
taken into account, the dark haloes are obtained much rounder, since
the baryons are more concentrated to the center, in the simulations of
Kazantzidis et al (2004).
These predictions are roughly compatible to the 
flattening observed statistically by 
weak lensing measurements (Hoekstra et al 2004), where the average 
ellipticity of the haloes is 0.33, if the dark matter is assumed aligned with the baryons.

\subsection{Results of various methods}

{\bf Method of HI flaring}

One of the methods most frequently  used is the flaring of the HI planes:
the thickness of the gaseous layers is a a function of the 
velocity dispersion, perpendicular to the plane and of the gravity restoring force
towards the plane, which depends on the mass contained within the HI layer.
The z-velocity dispersion is not known in the edge-on galaxies where the
HI thickness can be measured, but it is assumed the same as in similar face-on 
galaxies, where it is measured to be about constant with radius, and 
around 10km/s.  The thickness that can be derived by this method is
sensitive to the radial distribution of the dark matter and mainly to its 
extension or truncation at large radius, which is not known precisely.
Olling (1995, 1996), and Becquaert \& Combes (1997) found quite flattened
halos for NGC 224, 891, 3198, 4013 and 4244 (see Merrifield 2002).
From  an NFW profile of the dark matter halo in the outer parts 
(density in $r^{-3}$) Narayan et al (2005) find as the best fit 
a spherical shape. From the flaring also, 
Kalberla (2004) favors a model  where most of the 
dark matter is in the disk.

\bigskip

{\bf Polar rings}

As was shown above, the polar ring method is potentially
powerful, but many more systems should be studied in details. 
There is evidence for substantial dark matter, relatively flattened,
although  more along the polar plane than along the host plane.
Only the case of NGC 4650A is reported in Figure 3.

\bigskip

{\bf X-ray isophotes}

To determine the amount and the flattening of dark matter 
in elliptical galaxies, the method of the shape of X-ray isophotes 
have been used (e.g. Buote et al 1998, 2002). 
The method relies on the comparison
between the observed ellipticities and orientations of X-ray isophotes,
with those expected if the gravitating mass traces the same shape as the
stellar light. This method was applied to two isolated E4 ellipticals
(NGC 720 and NGC 3923), and one isolated S0 galaxy (NGC 1332).
The results were dark matter ellipticities between 0.4 and 0.6,
and the indication in NGC 720 that the dark matter is flatter than 
the stellar distribution (cf Fig 3).

\bigskip

{\bf Tidal streamers in the dark halo of the MW}

Most Sagittarius streams are too 
young to constrain the shape of the Milky Way dark halo.
Either oblate or prolate, with c=0.6 are compatible
with the data.
Only one older stream and its kinematics give a
prolate  shape of 0.6 Helmi (2004a,b). A more spherical fit is found
by Johnston et al (2005).

\begin{figure}[h]
   \centering
   \includegraphics[angle=-90,width=7cm]{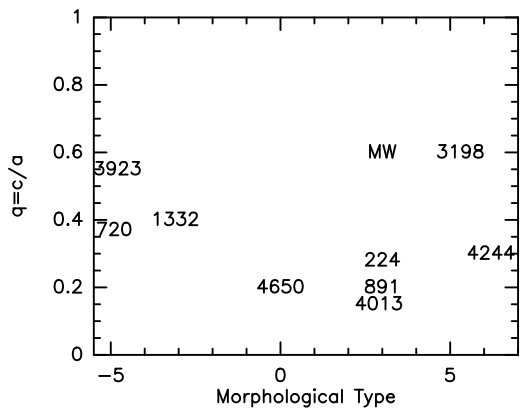}
      \caption{Distribution of total mass flattening (axis ratio q=c/a) as a function 
of morphological types, for the few galaxies studied in the literature, either by the polar ring,
X-ray isophotes or flaring HI plane techniques. Galaxies are marked by their NGC number, except
for the Milky Way. For the latter, the recent result from tidal streamers has been adopted.}
       \label{fig3}
   \end{figure}

\section{Warps and constraints on dark matter}

Warps in stellar disks are very frequent (e.g. Reshetnikov \& Combes 1998).
If the disk is self-gravitating, spontaneous bending instabilities
can explain these warps (Revaz \& Pfenniger 2004).
The disks are unstable to bending, if the z-velocity dispersion
is below Araki limit  $\sigma_z$/$\sigma_r$ = 0.293.
  If the optical disk warps have all to be explained through this
instability, then the fraction of dark matter in the disk
should be above a certain threshold, according to 
the flattening assumed for the spheroidal halo.
For a typical galaxy as the Milky Way, the dark matter should
be equally distributed in the disk and the 
spheroidal halo (Revaz \& Pfenniger 2004).

\section{Conclusions}

The kinematical observations of polar rings are precious tools to
probe the 3D shape of dark haloes. Their Tully-Fisher diagram
until now reveal that most PRGs have their dark matter 
flattened in the polar planes. More detailed data on more systems are
required to confirm this result, which would imply that
dark baryons could play a role in their formation.
  Small stellar warps that are present in most spiral disks put constraints
in the dark matter flattening, if it is assumed that
they are excited by bending waves. Large gas warps might be
a phenomenon akin to polar rings, formed through late accretion of
external gas. The frequency of these pehnomena confirm the 
important role of late gas accretion in galaxy formation.

%%-----------------------------
%%      your bibliography
%%-----------------------------

\end{document}